\documentstyle[aps,prl,multicol,epsf]{revtex}

\begin{document}
\def\CC{{\rm\kern.24em \vrule width.04em height1.46ex depth-.07ex
\kern-.30em C}}
\def\P{{\rm I\kern-.25em P}}
\def\RR{{\rm
         \vrule width.04em height1.58ex depth-.0ex
         \kern-.04em R}}

\newcommand{\be}{\begin{equation}}
\newcommand{\ee}{\end{equation}}
\newcommand{\bq}{\begin{eqnarray}}
\newcommand{\eq}{\end{eqnarray}}
\newcommand{\Sc}{Schr\"odinger\,\,}
\newcommand{\Sp}{\,\,\,\,\,\,\,\,\,\,\,\,\,}
\newcommand{\no}{\nonumber\\}
\newcommand{\tr}{\text{tr}}
\newcommand{\p}{\partial}
\newcommand{\la}{\lambda}
\newcommand{\La}{\Lambda}
\newcommand{\G}{{\cal G}}
\newcommand{\D}{{\cal D}}
\newcommand{\E}{{\cal E}}
\newcommand{\W}{{\bf W}}
\newcommand{\de}{\delta}
\newcommand{\al}{\alpha}
\newcommand{\bi}{\beta}
\newcommand{\ep}{\varepsilon}
\newcommand{\ga}{\gamma}
\newcommand{\epp}{\epsilon}
\newcommand{\vep}{\varepsilon}
\newcommand{\th}{\theta}
\newcommand{\om}{\omega}
\newcommand{\si}{\sigma}
\newcommand{\J}{{\cal J}}
\newcommand{\pr}{\prime}
\newcommand{\ka}{\kappa}
\newcommand{\TH}{\mbox{\boldmath${\theta}$}}
\newcommand{\DE}{\mbox{\boldmath${\delta}$}}
\newcommand{\lan}{\langle}
\newcommand{\ran}{\rangle}
\newcommand{\Hol}{\text{Hol}}
\newcommand{\cp}{{\bf CP}}

\draft
\title{ Non-Abelian Berry connections for quantum computation }
\author{Jiannis Pachos$^{1}$, Paolo Zanardi$^{1,2}$ and Mario Rasetti$^{1,3}$
}
\address{
$^{1}$ Institute for Scientific Interchange Foundation, \\
Villa Gualino, Viale Settimio Severo 65, I-10133 Torino, Italy\\
$^{2}$ Istituto Nazionale per la Fisica della Materia (INFM),
Unit\`a Politecnico di Torino \\
$^{3}$ Dipartimento di Fisica, Politecnico di Torino, Corso Duca, degli Abruzzi 24,
I-10129 Torino, Italy \\
}
\date{\today}
\maketitle
\begin{abstract}
In the holonomic approach to quantum computation
information is encoded in a degenerate eigenspace of a parametric family
of Hamiltonians and manipulated by the associated {\em holonomic gates}.
These are realized in terms of the non-abelian Berry connection and are obtained by 
driving the control parameters along adiabatic loops. 
We show how it is possible, for a specific model, to explicitly determine the loops
generating any desired logical gate, thus producing a universal set of unitary transformations.
In a multi-partite system unitary transformations can be implemented
efficiently by sequences of local holonomic gates.
Moreover a conceptual scheme for obtaining the required Hamiltonian family, 
based on frequently repeated pulses, is discussed, 
together with a possible process whereby the
initial state can be prepared and the final one can be measured.

\end{abstract}
\begin{multicols}{2}
\narrowtext
The field of quantum information and computation (QC) \cite{QC}
brings together ideas and techniques from very different areas ranging from
fundamental quantum physics to solid-state engineering 
and computer science. QC synergetically benefits from all these contributions
and conversely quite often offers fresh viewpoints on old subjects.
Recently it has been suggested \cite{ZARA} that
even tools related to gauge theories \cite{WY} might play a fruitful role
in the arena of QC. 
Indeed in ref. \cite{ZARA} 
the possibility of realizing quantum information processing
by using non-abelian Berry holonomies \cite{SHWI} induced by moving along suitable loops
in a control space $\cal M$ has been analysed.
The computational capability stems from
the features of the global geometry of the bundle of eigenspaces 
associated with a family $\cal F$ of Hamiltonians parametrized by points of $\cal M.$
The geometry is described by a non-trivial gauge potential $A$
or {\em connection},
with values in the algebra $u(n)$ of anti-hermitian matrices ($n$ is the dimension of the computational
space).
Since the unitary transformations realizing the 
 computations are nothing but the {\em holonomies}
associated with the connection A,
this conceptual framework for QC is referred to as Holonomic Quantum Computation (HQC).
In a sense HQC can be considered as the (continuous) differential-geometric
counterpart of the (discrete) topological QC with anyons described in refs. \cite{KIT,Preskill}.

In this paper we shall provide further analysis of this proposal.
After concisely reviewing the conceptual basis of HQC,
we shall show how, in a specific relevant model, one can explicitly determine the sequence of
loops necessary for generating any given 
quantum gate. Then we shall introduce HQC models with a natural 
multi-partite structure and discuss how this bears on the question of complexity.
Finally we shall discuss how in principle one can implement HQC 
by repeated pulse control of a system with degenerate spectrum.

Let us begin by recalling the basic ideas of HQC \cite{ZARA}.
Quantum information is encoded in a $n$-fold degenerate eigenspace $\cal C$ of 
a Hamiltonian $H_0$, with eigenvalue $ \varepsilon_0$.
Operator, $H_0$, belongs to 
a family ${\cal F}=\{H_\lambda\}_{\lambda\in{\cal M}}$, $H_0=H_{\lambda_0},$
in which
no energy level crossings occur as $\lambda$ ranges over $\cal M$.
In the following we shall satisfy this latter condition 
by assuming, for simplicity, 
that the Hamiltonians
$H_\lambda$ are isospectral ($H_\lambda = {\cal U}(\lambda)\,H_0\,{\cal U}(\la)^\dagger$).
The $\lambda$'s 
represent the ``control'' parameters that one has to drive in order
to manipulate the coding states $|\psi\rangle\in{\cal C}$.
In general the points of $\cal M,$ from the physical point of view,
can be thought of as describing external fields, 
such as electric or magnetic fields, or couplings between subsystems.
Let $C$ be a {\em loop} in the control manifold $\cal M,$ with base point $\lambda_0$
, $C\colon [0,\,1]\mapsto{\cal M},C(0)=C(1)=\lambda_0.$
We assume that $C$ is traveled along slowly with respect to the longest dynamical time scale involved:
in this case the evolution is adiabatic i.e., no transitions among different energy levels are
induced. 
If $|\psi\rangle_{in}\in{\cal C}$ is the initial state, at the end of this control process one gets
$ |\psi\rangle _{out}=e^{i\,\varepsilon_0\,T}\, \Gamma_{A}(C)
\,|\psi\rangle_{in}.$
The first factor here is just an overall dynamical phase and in the following it will be omitted;
let us just mention that such a decoupling of the fast dynamical 
evolution opens new possibilities about coherent and error avoiding encoding \cite{Preskill}.
The second contribution, the holonomy $\Gamma_{A}(C)\in U(n)$, 
has a purely geometric origin and its appearance
accounts for the non-triviality (curvature) of the {\em bundle} of eigenspaces 
over $\cal M.$
By introducing the Wilczek-Zee connection \cite{WIZE}
\begin{equation}
A^{\lambda_\mu}_{\bar \alpha \al}:= \langle\psi^{\bar \alpha}(\lambda)|
\,{\partial \over \partial\lambda_\mu}\,
|\psi_{}^\al(\lambda)\rangle ,
\label{conn}
\end{equation}
one finds
$\Gamma_{A}(C) ={\bf{P}}\exp \int_C A$ \cite{SHWI},
where ${\bf{P}}$ denotes path ordering.
The set Hol$(A):=\{\Gamma_{A}(C)\}_C \subset U(n)$ is known as the holonomy group
\cite{NAK}. In the case in which it coincides with the whole unitary
group $U(n)$ the connection $A$ is called {\em irreducible}.
In \cite{ZARA} it has been argued that for a large enough control manifold,
 the irreducible case is the ${\em generic}$
one, therefore one can in principle implement any computation over the {\em code} ${\cal C}$
just resorting to this very special class of quantum evolutions. 

{\em Quantum Gates.} 
A workable HQC model which represents a natural non-abelian
generalization of the original Berry phase, for which explicit
construction of the holonomic gates is possible, is now discussed.
The model is worked out with some details in that 
it is extendable to the more general case when $\cal M$ is 
a coset space.
The features of the construction presented are twofold. On the one hand 
it fully exploits the loop composition structure
at the basis of the holonomy group, showing a procedure whereby loops can be 
decomposed into $2$-dimensional components, simple to deal with.
On the other this topological construction overcomes 
the difficulties connected with the path ordering prescription.

Let us consider the Hamiltonian $H_0=\ep_0 | n+1\rangle \langle n+1 |$
acting on the state-space ${\cal H}\cong \CC^{n+1}=\mbox{span} \{|\al \ran \}_{\alpha=1}^{n+1}.$
We shall take as the family ${\cal F}$ the whole orbit 
${\cal O}(H_0):=\{ {\cal U}\,H_0\,{\cal U}^\dagger\,/\,{\cal U}\in U(n+1)\}$ of $H_0$
under the (adjoint) action of the unitary group $U(n+1).$ 
This orbit is isomorphic to the $n$-dimensional complex projective space:
\begin{eqnarray}
{\cal O}(H_0)
\cong \frac{ U(n+1)}{U(n)\times U(1)} \cong\frac{SU(n+1)}{U(n)}
\cong
{\bf{CP}}^n.
\nonumber
\end{eqnarray}
The points of ${\bf{CP}}^n$ can be parametrized by the unitary matrices
 ${\cal U}(z) =U_1(z_1)U_2(z_2)...U_n(z_n)$ where
$U_\al(z_\al)=\exp [ G_\al(z_\al)]$ with 
$G_\al(z_\al)=z_\al |\al\ran\lan n + 1| -\bar z _\al |n ~+~ 1 \ran \lan \al |$, 
 $z_\al=\th_\al e^{i \phi_\al}.$ 
The eigenstates of the rotated Hamiltonians are 
$
|\al(\th,\phi)\ran:={\cal U}(\th, \phi) |\al \ran =\cos \th_\al |\al \ran -\exp ({-i \phi_\al})$ $
\sin \th_\al \sum_{j>\al} ^{n+1} \exp({i \phi_j}) \sin \th_j$ $ \prod_{j>\ga>\al} \cos
\th_\ga |j \ran
$
and 
$
|n+1(\th, \phi) \ran :={\cal U}(\th, \phi) |n+1 \ran =\sum^{n+1}_{j=1} \exp ({i \phi_j}) \sin
\th_j \prod_{\ga<j} \cos \th_\ga |j \ran 
$,
where $\th_{n+1}:=\pi/2$ and $\phi_{n+1}:=0$.
Notice that for $n=1$ the standard $2$-level model with (abelian) Berry phase is recovered.
By using equation (\ref{conn}) the components of the connection can be now
explicitly computed. The only non-zero elements of the matrix $A^{\th_\bi}$ ($\beta=1,\ldots,n$)
are
$A^{\th_\bi}_{\bar \al \bi}$ for $\bar \al=1,\ldots, \bi-1$, given by
\begin{eqnarray}
A^{\th_\bi}_{\bar \al \bi}=e^{i(\phi_{\bar \al}- \phi_\bi)}\, \sin \th_{\bar \al} 
\prod_{\bi>\ga>\bar \al} \cos \th_\ga \,\,,
\label{niao}
\end{eqnarray}
as well as $A^{\th_\bi}_{\bar \al \bi}=-A^{\th_\bi}_{\beta \bar\alpha}.$
The antihermitian matrix $A^{\phi_\bi}$ has non-zero elements for 
$\al=\bi$ and $\al \geq \bar \al$ given by
\begin{eqnarray}
A^{\phi_\bi}_{\bar \al \bi}=-i e^{i(\phi_{\bar \al}-\phi _\bi)} \sin
\th_\bi \sin \th_{\bar \al} \prod _{\bi \geq\ga > \bar \al } \cos \th_\ga,
\nonumber
\end{eqnarray}
with $\prod _{\bi \geq\ga > \bi } \cos \th_\ga=1$, 
and for $\bi>\al$ and $\al\geq \bar \al$ by
\begin{eqnarray}
A^{\phi_\bi}_{\bar \al \al} &=& i e^{i(\phi_{\bar \al}-\phi _\al)} \sin \th_{\al} \sin
\th_{\bar \al} \sin^2 \th_\bi
\!\!\! \prod _{\bi >\ga > \al }\!\!\!\! \cos \th_\ga \! \! \!
\!\prod _{\bi > \bar \ga >\bar \al } \!\! \! \! \cos \th_{\bar \ga}.
\nonumber
\end{eqnarray}
The $A^{\th_\bi}$'s and $A^{\phi_\bi}$'s are the $2\,n$ components
of the $u(n)$-valued connection over ${\bf{CP}}^n.$ 

For generating a given quantum gate $g\in U(n)$
one has to determine a loop $C_g$ in $\cal M$ such that $\Gamma_A(C_g)=g$. 
Due to the non-abelian character of the connection such an inverse problem
is in general hard to solve. To tackle it we choose specific families of
loops $\{C_i\}$, that generate holonomies from which one can eventually construct any $U(n)$
transformation. To this end we consider the $2$-dimensional submanifolds in the $2\, n$-dimensional
space $(\th, \phi)$, spanned by two variables, $(\th_\bi,\phi_{\bar
\bi})$ or $(\th_\bi, \th_{\bar\bi})$, for specific values of $\bi$ and $\bar \bi$. 
For these loops the line integral is given by
$\oint _C A = \oint_C ( A^{\th_\bi} d\th_\bi +A^{ \la_{\bar \bi}} d\la_{\bar \bi })$,
where $\la =\th$ or $\phi$. From (\ref{niao}) we see that
we can always choose the parameters which define the position of the plane $(\th_\bi , \la_{\bar \bi})$, 
where loop $C$ lies, in
such a way that the matrix $A^{\th_\bi}$ is identically zero. 
If one takes $\th_i=0$, $\forall i\neq \bi, \,\, \bar \bi$, 
matrices $A^{\th_\bi}$ and $A^{\la_{\bar \bi}}$ commute, so that we can calculate the 
integral and exponentiate avoiding the path ordering problem.

In this framework, it is possible to identify first 
 four families of loops in such a way as to produce the basis of
four matrices (the Pauli matrices and the identity) of all possible two-by-two 
submatrices belonging in the algebra of $U(2)$. The first choice is $(\th_\bi, \phi_\bi)$, where
the non-zero component of the connection is $A^{\phi_\bi}_{\bi \bi}=-i \sin^2 \th_\bi$.
The second choice is the loop on submanifold $(\th_\bi, \phi_{\bar \bi})$ for $\bar \bi > \bi$, 
with $\th_{\bar \bi}=\pi/2,$ giving a
different connection with two non-zero elements, 
$A^{\phi_{\bar \bi}} _{\bi \bi}=i \sin^2 \th_\bi$ and 
$A^{\phi_{\bar \bi}} _{\bar \bi \bar \bi}=-i$. Of course the latter element will give zero
when integrated along a loop. 
For $\bar \bi < \bi$ both matrices are
identically zero, and give rise to trivial holonomy. With these two connections
and for appropriate loops one can
 obtain all possible $U(n)$ {\it diagonal} transformations.
For loop
$
C_1 \in (\th_\bi, \phi_\bi),$
$\Gamma_A(C_1)=
\exp [ -i |\bi \ran \lan \bi | \Sigma_1 ],
$
$\Sigma_1$ denoting the
area enclosed by $C_1,$ on the $S^2$ sphere with coordinates $(2\th_\bi,\phi_\bi)$. 
For 
$C_2 \in (\th_\bi, \phi_{\bar \bi}),$ 
$\Gamma_A(C_2)=
\exp [i |\bar \bi \ran \lan \bar \bi | \Sigma_2]
.$
Recalling the constraint $\bar \bi> \bi$, we see that one can produce $n-1$ distinct holonomies
from $C_2$.

To obtain the non-diagonal transformations one has to consider a loop on the $(\th_\bi,
\th_{\bar \bi})$ plane, with $\th_i=0$ for all $i\neq \bi,\bar \bi$. Then the only non-vanishing
elements of the connection are $A^{\th_{\bar \bi}}_{\bi \bar \bi}=e^{i(\phi_{\bi}-\phi_
{\bar \bi})} \sin \th_\bi= -\bar A^{\th_{\bar \bi}}_{\bar \bi \bi}.$
By choosing further the 
$(\theta_\beta,\,\th_{\bar \bi})$ plane at $\phi_\bi=\phi_{\bar \bi}=0$ the 
holonomy becomes,
for 
$C_3 \in (\th_\bi, \th_{\bar \bi})_{\phi_\bi=\phi_{\bar
\bi}=0},$ 
$\Gamma_A(C_3)= 
\exp [ -(|\bi \ran \lan \bar \bi| - |\bar \bi \ran \lan \bi |)\Sigma_3 ],
$
while at $\phi_\bi=\pi /2$ and $\phi_{\bar \bi}=0,$ 
for 
$C_4 \in (\th_\bi, \th_{\bar \bi})_{\phi_\bi=\pi/2,\phi_{\bar
\bi}=0},$ 
$\Gamma_A(C_4)= 
\exp [-i (| \bi \ran \lan \bar \bi| + |\bar \bi \ran \lan \bi |) \tilde \Sigma _4 ] ,
$
where $\tilde\Sigma$ is the area on the sphere with coordinates 
$(\pi/2 -\th_\bi, \th_{\bar \bi})$. Note that any loop $C$ on the $(\th_\bi, \la_{\bar \bi})$
plane with the same enclosed area (when mapped on the appropriate sphere) 
$\Sigma_C$ will give the same holonomy independent of its position and shape. These four
holonomies restricted each time to a specific $2\times2$ submatrix, generate all
$U(2)$ transformations. Finally it is easy to check that in this way one can indeed obtain
$U=\exp[\mu_j\, T_j]$, where $T_j\,(j=1,\ldots,n^2)$ is a $u(n)$ generator
and $\mu_j$ an arbitrary real number.
Therefore any element of $U(n)$ can be obtained
by controlling the $2\,n$ parameters labelling the points of ${\bf{CP}}^n.$

It is instructive to consider the form that the Hamiltonian family $\cal
F$, takes when restricted to the particular 2-submanifolds. 
For the loop $C_1$ (similarly for $C_2$) one finds $H_1=-\ep_0/2\,
\vec{B}(2\th_{\bi},\phi_{\bi})\cdot\vec{\sigma}$ for
$\vec{B}(\th_i,\phi_j) =(\sin \th_i \cos \phi_j, \sin \th_i \sin \phi_j,
\cos \th_i)^T$, where the only non-zero elements are on the
$\bi$-th and $(n+1)$-th row and column. $H_1$ generates an abelian ${\bf CP}^1$
phase between the states $|\bi\ran$ and $|n+1 \ran$. 
On the other hand for the path $C_3$ (similarly for $C_4$) we have $H_3=\ep_0
{\vec B}(\th_\bi,\th_{\bar \bi}) {\vec B}(\th_\bi,\th_{\bar \bi})^T$, where the
non-zero elements connect the states $|\bi\ran$, $|\bar \bi \ran$
and $|n+1\ran$. In this 
Hamiltonian there is direct coupling between three states, giving rise to a
non-abelian interaction.

In order to represent a $2$-qubit system we have to consider the control manifold $\cp^4$.
The holonomies in this case are $4\times4$ matrices, and we take as a representation basis of the 
unitary transformations the qubit basis $|00\ran$, $|01\ran$, $|10\ran$ and $|11\ran$. 
From the general scheme above it follows that by appropriate control of the
parameters $(\th, \phi)$ for obtaining various loops $C$, we can generate all
possible $U(4)$ rotations i.e., any logical gate, 
 in particular 
single-qubit rotations, and two-qubit gates such as the controlled operations XOR and CROT. 
($U_{XOR}:=|0\rangle\langle0|\otimes\openone+ |1\rangle\langle 1|\otimes \sigma_x,\,
U_{CROT}:=|0\rangle\langle0|\otimes\openone+ |1\rangle\langle 1|\otimes \sigma_z$).
For single qubit rotations we consider three unitaries $U_\alpha=\Gamma_A(C_\alpha),\,(\alpha=1,2,3)$ with 
$C_1\in(\th_1,\phi_1)_{\th_2=\phi_2=0},$ $C_2\in(\th_1,\phi_2)_{\th_2=\pi/2,\phi_1=0},$
$C_3\in
(\th_1,\th_2)_{\phi_1=\phi_2=0}.$
For areas $\Sigma_2=\Sigma_1$ let $U_{ph1}=(U_1U_2) U_3(U_1U_2)^{-1}$, equal to
\be
U_{ph1}=
\left[ \begin{array}{cccc}   
\begin{array}{ccc}         \cos \Sigma_3 & -\sin \Sigma_3 e^{-2i\Sigma_1}\\
                           \sin \Sigma_3 e^{2i\Sigma_1} & \cos \Sigma_3 \\
\end{array} & {\bf 0} \\
 {\bf 0}  & {\openone} \\
\end{array} \right]. 
\ee
With loops on the $(\th_3,\phi_3)_{\th_4=\phi_4=0}$, $(\th_3,\phi_4)_{\th_4=\pi/2,
\phi_3=0}$, $(\th_3,\th_4)_{\phi_3=\phi_4=0}$ planes and spanning the same areas as
before, one can produce the rotation $U_{ph2}$ on the lower-right block,
and with their product obtain the phase rotation of one qubit ${\openone} \otimes
U_{q}=U_{ph1}U_{ph2}$. 
In order to perform the phase rotation of the other qubit all we need is 
the swapping operator, $S$, acting on the two qubits 
by $S\,|\psi\ran\otimes|\phi\ran=|\phi\ran\otimes |\psi\ran$.
$S$ is given in terms of holonomies, by $S=$ $\left.\Gamma_A(C_4)\right|_{\tilde 
\Sigma_4={\pi/ 2}} \left.\Gamma_A(C_2)\right|_{\Sigma_2={\pi/ 2}}$,
where $C_4\in(\th_2,\th_3)_{\phi_2=\pi/2,\phi_3=0}$ and $C_2\in(\th_1,\phi_2)_{\th_2=\pi/2}$ and
$(\th_1,\phi_3)_{\th_3=\pi/2}$.
Hence, the phase rotation of the other qubit is given by $U_{q}\otimes 
{\openone}=S({\openone} \otimes U_{q})S$. The controlled rotation gate CROT, 
is given by $U_{CROT}=\left.\Gamma_A(C_1)\right|_{\Sigma_1=\pi/2}
\left. \Gamma_A(C_2)\right|_{\Sigma_2=\pi/2}$, 
where $C_1 \in (\th_4,\phi_4)$ and $C_2\in (\th_1,\phi_4)_{\th_4=\pi/2}$.
In addition the `exclusive or' gate XOR 
can be realized as $U_{XOR}=\left.\Gamma_A(C_4)\right|_{\tilde\Sigma_4=\pi/2}
\left.\Gamma_A(C_2)\right|_{\Sigma_2=\pi/2}$, 
where $C_4 \in (\th_3,\th_4)_{\phi_3=\pi/2, \phi_4=0}$ and $C_2 \in (\th_1,\phi_3)_{\th_3=\pi/2}$ and 
$(\th_1,\phi_4)_{\th_4=\pi/2}$.

{\em Complexity.}
So far the coding subspaces analysed for HQC do not necessarily involve quantum entanglement;
${\cal H}$ could be the state-space of a {\em single} quantum system.
This is due to the fact that $\cal C$ does not have a built-in tensor product structure,
thus, in general, it cannot be naturally interpreted as the state-space of a multi partite
system. 
This latter feature, however, is one of the essential ingredients that make QC more efficient than 
classical computation.
Indeed, from the above construction for the ${\bf{CP}}^n$-model it is easy to realize
that the number of elementary loops in $2$-submanifolds necessary for implementing
single and two-qubit operations scales exponentially as a function of the number $\log_2 n$
of encoded qubits.
To overcome such a difficulty one has simply to consider an 
Hamiltonian family, acting on the state-space of a multi-partite quantum system
[not just isomorphic to], with a special structure allowing for local QC's to be holonomically performed.
Then global QC's involving non-trivial actions over many qubits, can be efficiently realized
in the standard way \cite{HOW}. 
One possible formalization of this idea is contained in the following
theorem:

Let ${\cal H}:=\otimes_{j=1}^{n} {\cal H}_j\otimes{\cal H}_a,
\,{\cal H}_j\cong{\bf{C}}^d,$ and $ {\cal H}_a=\mbox{span}\{|-\rangle,\,|+\rangle\}$
 a single-qubit ancillary space.
We set 
$H_a:= \varepsilon\,\sigma^z \in\mbox{End}({\cal H}_a)$ \cite{not}.
Let moreover $H(\lambda):=\sum_{i<j} V_{ij}(\lambda_{ij}),$ where the $\lambda_{ij}$'s 
belong to local control manifolds ${\cal M}_{ij}.$ Suppose that
$V_{ij}(\lambda)\in\mbox{End}({\cal H}_i\otimes {\cal H}_j\otimes{\cal H}_a),$
and $V_{ij}(0)=H_a+ H_i+H_j,$ 
where $H_j\in\mbox{End}({\cal H}_j)$ is such that $H_j\,|\alpha\rangle_j=0,\,(\alpha=0,1)$
and that
the family
$\{V_{ij}(\lambda)\}_{\lambda\in{\cal M}_{ij}}$
allows for universal HQC over the degenerate eigenspaces 
$
{\cal C}_{ij}^\pm :=\mbox{span}\,\{|\alpha\rangle_i\otimes |\beta\rangle_j \otimes |\pm\rangle\,\colon\,\alpha,\beta=0,1\}
\cong{\bf{C}}^2\otimes{\bf{C}}^2.
$
Then
the family $\{H(\lambda)\,/\,\lambda\in\prod_{i<j} {\cal M}_{ij}\}$ allows for {\em efficient} 
universal HQC
over the $n$-qubit codes ${\cal C}^\pm:=\mbox{span}\{\otimes_{j=1}^n|\alpha_j\rangle\otimes|
\pm\rangle\,/\,\alpha_j=0,1\} \cong ({\bf{C}}^2)^{\otimes\,n}.$

The proof of the latter proposition proceeds
as follows.
First one observes that the subspaces ${\cal C}_{ij}^\pm$ are degenerate eigenspaces of $V_{ij}(0)$
corresponding
to the eigenvalues $\pm\varepsilon.$ Hence the ${\cal C}^\pm$'s are $2^n$-dimensional eigenspaces of $H(0)$
with eigenvalues $\pm n\,(n-1)\,\varepsilon/2.$
By assumption for any pair $(i,\,j)$ of subsystems \cite{local}
 one can generate any unitary transformation over
${\cal C}_{ij}^\pm$ by adiabatic loops $\gamma_{ij}$ in ${\cal M}_{ij}.$ Keeping all the remaining $\lambda$'s
at $0,$ 
one has a trivial action over the other factors of ${\cal C}^\pm,$
 while
$
\Gamma_A(\gamma_{ij})\colon \otimes_k |\alpha_k\rangle\otimes|\pm\rangle\mapsto \sum_{\beta_i \beta_j} 
U_{\beta_i \beta_j,\,\alpha_i \alpha_j}(\gamma_{ij}) \otimes_k |\tilde \alpha_k\rangle\otimes|\pm\rangle
,$
where $\tilde \alpha_k=\alpha_k$ for $k\neq i,j$ and 
$\tilde \alpha_k=\beta_k$ when $k=i,j.$
In particular one can obtain a universal set of gates e.g., XOR$^\pm_{ij},$ and single-qubit
operations, by using a fixed amount of resources. 
The claim then follows by well-known universality results for QC \cite{UG}.
The above scheme involves the use of an ancilla and requires
controllability of three-body interactions, extremely difficult to achieve
in practice. In this respect a simplification, involving just two-body interactions, can be obtained
by considering $N$ subsystems with $d$-levels \cite{trits} .

{\em Implementation.}
Now we discuss how one could in principle implement the holonomic loops
even when the parametric Hamiltonian family $\cal F$ (or a part of it)  is not available
from the outset.
We shall resort to ideas of quantum control theory in a way quite similar
to the one adopted for symmetrization procedures \cite{SYMM} and decoherence control in open quantum systems
\cite{VIO}.
Suppose that an experimenter has at disposal 
the following resources:
i) A quantum system characterized by the Hamiltonian $H_0$
admitting a $n$-fold degenerate eigenspace ${\cal C};$
ii) The way to turn on and off a set of interactions
very quickly (with respect to the time-scales associated with $H_0$)
in such a way that a set of unitary ``kicks'' 
$K:=\{U_\lambda\}_{\lambda\in{\cal M}}$ 
can be realized.
Let $T=N\,\Delta\,t$ and $t_0=0,\,t_{i+1}=t_i+\Delta\,t\,(i=1,\ldots\,N-1)$
be a partition of the time interval $[0,\,T].$
Now let the system evolution to be as follows:
at any time $t_i$ the experimenter kicks the system with the pulse
$U^\dagger_{i+1}\,U_{i}\,$ where $U_i:=U(\lambda_i)$ is a
 unitary chosen from the set $K$ ($U_0=U_N=\openone$).
Between the kicks the system evolution is unperturbed,
$U(\Delta t)=e^{-i\,H_0\,\Delta t}.$
The global evolution is then given by
\begin{equation}
U_{N,\Delta t} (T)={\bf {T}}\,\prod_{i=1}^{N-1}U_i\,U(\Delta t)
\,U_i^\dagger ,
\nonumber
\end{equation}
where ${\bf {T}}$ denotes time-ordering.
By considering the limit $\Delta\,T\mapsto 0,\,N\mapsto\infty,$
with $N\,\Delta t=T$ 
one gets 
$ U_{N,\Delta t} (T) \rightarrow {\bf {T}} \exp\,-i\int_0^T dt\, H(t) 
$
where $H(t):= U(\lambda(t))\,H_0\,U(\lambda(t))^\dagger.$
In particular by making the function $\lambda(t)$ vary adiabatically,
 one can obtain the desired holonomic evolution. 
This scheme is based on a strong separation between time-scales:
each $U_\lambda$ has to be enacted impulsively whereas
the characteristic variation time of the control parameters $\lambda$ has to be
slow enough to satisfy the adiabaticity requirement. 
More precisely,
if $\tau_k$ denotes the kicking time, $\omega$ the (highest)
frequency associated to the dynamics generated by $H_0$
and $\tau_\lambda$ the time-scale over which the function $\lambda(t)$ varies,
 one must have
$\tau_k\le \Delta t\ll\omega^{-1}\ll \tau_\lambda.$
 Notice that
the pulses $U_\lambda,$ are not required to be a universal set of gates for  QC;
here they represent an extra resource needed for implementing HQC
when $\cal F$ is missing.

Finally let us briefly consider the problem of codeword preparation and measurement.
In order to encode the initial state into the degenerate subspace $\cal C$,
or to make the measurement on the final state,
it would be useful to lift the degeneracy.
Indeed in this way one would be able to distinguish
{\em energetically}
the different coding states. This characteristic
is quite often desirable from the experimental point of view
in that one can resort to procedures involving energy transitions with state-dependent 
frequencies.
The idea is to lift the degeneracy between the coding states, for
example by switching on coherently an external (generic) perturbation.
The basis states $|\psi_\alpha\rangle$ of $\cal C$ 
are mapped onto a set of states $|\psi^\prime_\alpha\rangle$
that are no longer energy degenerate. Preparation/measurement are then performed and 
eventually degeneracy is coherently restored by switching off the perturbation.

{\em Summary}.
In this paper we have provided further analysis
of the proposal for Holonomic Quantum Computation of ref. \cite{ZARA}.
We have explicitly designed control-loops whose holonomies 
generate universal gates for a $\cp^n$ control parameter manifold.
The basic idea is to associate to the $U(n)$-generators computable transformations
obtained by loops on $2$-dimensional subspaces of the control
manifold. 
Explicit realization of two qubit gates has been given with the indication of which 
particular loops the experimenter has to perform. 
In terms of such elementary holonomic gates
 we analysed the complexity problem and 
we showed how to achieve efficient implementation of
quantum computing by resorting to a HQC model
involving only local interactions.
Some implementative issues have been addressed, and
we devised a scheme based on repeated pulses
for realizing the parametric 
family of isospectral Hamiltonians required for HQC.
Finally, we briefly indicated how to prepare the initial
state and how to measure the final one
by coherently switching on and off the energetic degeneracy
of the computational subspace.

J.P. acknowledges a TMR Network support under the contract no. ERBFMRXCT96-0087.
P.Z. is supported by Elsag, a Finmeccanica company.

\end{multicols}


\begin{references}
\bibitem{not} The notation
$X\in\mbox{End}({\cal H}_j)$ means that $X$ has non-trivial action only on the $j$-th factor of ${\cal H}$.
\bibitem{local}
When the system   geometry
is given,
one can restrict  to ``local'' gates 
without  loss of efficiency.
\end{references}
\end{document}